\documentclass[12pt]{article}

\pdfoutput=1

\usepackage{cite,color}
\usepackage{mathrsfs,amsmath,bm,fixmath}

\catcode`@=11

\let\@CITE=\@cite
\let\Cite=\cite
\def\@cite#1#2{{#1\if@tempswa , #2\fi}}

\def\cite#1{[\Cite{#1}]}

\catcode`@=12

\evensidemargin=\oddsidemargin
\def\Eqn#1{Eq.\ (\ref{#1})}
\def\Eqs#1#2{Eqs.\ (\ref{#1}) and (\ref{#2})}
\def\next{\nonumber\\}
\def\sect#1{Sec.\,\ref{#1}}
\def\textcite#1{Ref.\ \cite{#1}}

\def\lag{{\mathscr L}}
\def\tilde{\widetilde}
\def\sub#1{_{\rm #1}}
\def\del{\partial}

\title{\bf Covariant formulation of electrodynamics \\ 
  in isotropic media}

\author{\bf Palash B. Pal \\
Physics Department, University of Calcutta,\\ 92 APC Road, Calcutta
700009, India}

\date{}

\begin{document}

\maketitle

\begin{abstract}

  The equations of electromagnetic fields in a medium is usually
  written in the rest frame of the medium.  We outline a method of
  generalizing the discussion to arbitrary inertial frames.  In the
  discussion, we also include the possibility that the medium is
  optically active, a possibility that is often overlooked in
  discussions of electromagnetic fields in a medium.
  
\end{abstract}

\bigskip

\section{Aim of the paper}\label{s:aim}
Covariant formulation of electrodynamics in the vacuum is a subject
that appears in standard textbooks of electrodynamics
\cite{textbooks}.   Electrodynamics in material media is also a
standard subject, but is usually not discussed in a manifestly
covariant manner.  A medium of course provides a preferred frame for
the discussion, and this frame is used for the standard formulations.
However, a preferred frame does not preclude us from discussing the
subject from other frames.  When we discuss the decay of a particle,
for example, there is a preferred frame, viz.\ the rest frame of the
particle; but it does not mean that we cannot ask what would be the
lifetime in any other frame.  Similarly, for the case of
electrodynamics, we might be interested about problems with a moving
medium, and it is mandatory to obtain a Lorentz-covariant formulation
of electrodynamics in a medium in order to discuss such subjects.

There have been attempts \cite{Hehl:2016wwp, Thompson:2017tus} at
general co-ordinate invariance, something that would be consistent
with the general theory of relativity.  This is a considerably
involved topic, and will not be touched on in this paper.  We keep our
attention on a formulation consistent with the special theory of
relativity, i.e., a Lorentz-invariant formulation.  Surely, there are
previous attempts in this direction as well.  Some of them have been
aimed at a quantum field theoretic formulations of the problem
\cite{Kislinger:1975uy, Nieves:1989uh}.  They usually employ the
scalar and vector potentials, encapsulated in the 4-vector potential
$A^\mu$, to formulate the problem.  There are also other attempts
\cite{Starke:2017wri} where the response equations do not always look
manifestly Lorentz-invariant, although covariance can be proved with a
certain amount of effort.

Our aim in this article is twofold.  First, we note that the presence
of the medium implies an extra 4-vector in the problem, viz.\ the
velocity 4-vector of the center of mass of the medium, $u^\mu$.  While
this 4-vector has been extensively used in the quantum field theoretic
formulations of electrodynamics in a medium \cite{Kislinger:1975uy,
  Nieves:1989uh, Weldon:1982aq}, to our knowledge it has not been used
in textbook-level formulation of classical electrodynamics.  We use
only gauge-invariant objects in the discussion, except where $A^\mu$
is absolutely indispensable --- viz., in the Lagrangian formulation,
for writing the interaction of the field with the sources of the
field.

Secondly, we include the description of natural optical activity in
our formalism.  Hardly any textbook on electrodynamics discusses this
phenomenon.  In one exception that does \cite{LL}, tensorial responses
in the medium seems essential to the explanation.  However, even
isotropic systems (like a sugar solution) show natural optical
activity, so tensorial response functions should not be essential for
describing the phenomenon.  In the context of quantum field
theoretical formulation, it was shown \cite{Nieves:1988qz} that the
most general linear response of an isotropic medium contains three
response functions, i.e., one more above the dielectric function and
the magnetic permeability.  Later, the need for this extra function
was demonstrated in the classical formulation, using 3-dimensional
vectors for the electric and magnetic fields \cite{Nieves:1992et}.
Here, we show how this extra constant can be accommodated in a
completely covariant formulation.

Throughout, we use the Heaviside-Lorenz system of electromagnetic
units, which is the most suitable system for a relativistic
formulation.

\section{Summary of covariant electrodynamics in the vacuum}\label{s:vac}
We assume that the reader is familiar with the 4-dimensional covariant
formulation of classical electrodynamics in the vacuum.  We are
presenting the key results here \cite{textbooks} in order to set up
the notation and to aid the discussion of subsequent sections.

In the covariant formulation, the electromagnetic field is represented
by a rank-2 antisymmetric tensor $F_{\mu\nu}$, where the Greek indices
are spacetime indices, 0 being the temporal direction and 1,2,3 the
spatial directions.  This tensor can be written as
\begin{eqnarray}
  F_{\mu\nu} = \del_\mu A_\nu - \del_\mu A_\nu \,,
  \label{F}
\end{eqnarray}
where $A_\mu$ is the 4-vector for electromagnetic potentials, and
\begin{eqnarray}
  \del_\mu \equiv {\del \over \del x^\mu} \,.
\end{eqnarray}
The sources of the electromagnetic field are summarized by a 4-vector
$J^\mu$ that is a function of position and time, just as $F_{\mu\nu}$
is.  The relation is through the differential equation which can be
written as
\begin{subequations}
  \label{maxwell4d}
\begin{eqnarray}
  \del_\mu F^{\mu\nu} = \frac1c J^\nu
  \label{dF}
\end{eqnarray}
by suitably adjusting a  constant factor in the definition for
$J^\mu$.  Written in 3-dimensional vector notation, this equation
contains two of the Maxwell equations, in particular the inhomogeneous
ones.  The other two, i.e., the homogeneous equations, can be written
in the 4-dimensional notation as
\begin{eqnarray}
  \del_\mu \tilde F^{\mu\nu} = 0 \,,
  \label{dFtil}
\end{eqnarray}
\end{subequations}
where $\tilde F^{\mu\nu}$ is called the dual of the field tensor
$F_{\mu\nu}$, defined as
\begin{eqnarray}
  \tilde F^{\mu\nu} = \frac12 \varepsilon^{\mu\nu\lambda\rho}
  F_{\lambda\rho} 
  \label{Ftil}
\end{eqnarray}
with the help of the completely antisymmetric Levi-Civita symbol in
4-dimensional spacetime.  While \Eqn{dFtil} follows from the
definitions in \Eqs{F}{Ftil}, the inhomogeneous equations of \Eqn{dF}
can be derived from a Lagrangian density
\begin{eqnarray}
  \lag = - \frac14 F_{\mu\nu} F^{\mu\nu} - \frac1c J^\mu A_\mu \,.
  \label{lagF}
\end{eqnarray}
The Euler-Lagrange equation, treating the $A^\mu$'s as the fields, are
\begin{eqnarray}
  \del_\alpha \left( {\del \lag \over \del (\del_\alpha A_\beta)}
  \right) = {\del \lag \over \del A_\beta} \,.
\end{eqnarray}
When one uses the derivative rules
\begin{subequations}
  \label{derivs}
\begin{eqnarray}
  {\del (\del_\mu A_\nu) \over \del (\del_\alpha A_\beta)} &=&
  \delta^\alpha_\mu \delta^\beta_\nu \,,
  \label{ddA/ddA} \\ 
  {\del A_\nu \over \del A_\beta} &=& \delta^\beta_\nu \,, 
  \label{dA/dA}
\end{eqnarray}
\end{subequations}
one obtains \Eqn{dF} in a straightforward manner.

\section{Medium with a linear response}\label{s:lin}
A medium contains charged particles.  Thus, in any electromagnetic
problem in a medium, there are two kinds of sources of the
electromagnetic fields.  If all these charges and currents, along with
externally placed sources, are taken into account, the equations given
in \sect{s:vac} are still valid.  However, they are not convenient, or
maybe even impossible, to use in practical problems, because of the
difficulty of accounting for all charges and currents of the particles
that constitute in the medium itself.  The usual escape route is to
parametrize the charges and currents bound within the medium by some
quantities and set up the equations with the free charges and
currents, over which an experimenter can have any kind of handle.  We
define the 3-vectors $\vec D$ and $\vec H$ whose sources are the free
charges and currents, and write the inhomogeneous Maxwell equation in
terms of them.  These vectors are assumed to be linearly related to
the electric and magnetic fields, $\vec E$ and $\vec B$:
\begin{eqnarray}
  \vec D = \epsilon \vec E \,, \qquad \vec H = \frac1\mu \vec B \,.
  \label{DH3}
\end{eqnarray}
We will assume these linear relations throughout this article.

As explained in \sect{s:aim}, we need a covariant formulation for
discussing problems with a moving medium.  The problem is that, the
coveriant formulation in the vacuum uses the tensor $F_{\mu\nu}$, and
we cannot say that in a medium we have an antisymmetric tensor that is
proportional to $F_{\mu\nu}$ and which depends only on the free
charges and currents.  In other words, we cannot simply write an
equation
\begin{eqnarray}
  \del_\mu G^{\mu\nu} = \frac1c J^\nu \sub{free} \,,
  \label{dG}
\end{eqnarray}
and claim that $G^{\mu\nu}$ is proportional to $F^{\mu\nu}$, because
that way one would obtain just one parameter connecting the two
tensors, whereas \Eqn{DH3} contains two parameters of proportionality,
or two response functions.

However, there is at least another 4-vector connected to the physical
description of the problem.  We are talking about a medium which can
be in motion in the frame that we choose.  If the medium, as a whole,
moves with a velocity $\vec v$ in the frame of reference, we can
define its dimensionless velocity 4-vector,
\begin{eqnarray}
  u^\mu = {1 \over \sqrt{1 - v^2/c^2}} \Big\{ 1, \vec v/c \Big\} \,,
\end{eqnarray}
which satisfies the relation
\begin{eqnarray}
  u^\mu u_\mu = 1 \,.
  \label{uu}
\end{eqnarray}
We can use this vector to formulate the equations of electrodynamics
in a medium.

Immediately, we notice that we can use this 4-vector and the
electromagnetic field tensor to define two new
4-vectors~\cite{Nieves:1989uh, Padmanabhan:2008gr}: 
\begin{subequations}
  \label{EB}
\begin{eqnarray}
  E^\mu &=& F^{\mu\nu} u_\nu \,, \\
  B^\mu &=& - \tilde F^{\mu\nu} u_\nu = - {1 \over 2}
  \varepsilon^{\mu\nu\lambda\rho} u_\nu F_{\lambda\rho} \,.
\end{eqnarray}
\end{subequations}
We have adjusted the signs in these equations in such a way that, in
the rest frame of the medium, where the only non-zero component of
$u^\mu$ is the time component, the components of these two 4-vectors
are given by
\begin{subequations}
  \label{EB0}
\begin{eqnarray}
  E^\mu &\xrightarrow{\quad v=0 \quad} & \Big\{ 0, \vec E \Big\} \,,
  \label{E0}\\ 
  B^\mu &\xrightarrow{\quad v=0\quad} & \Big\{ 0, \vec B \Big\} \,,
  \label{B0}
\end{eqnarray}
\end{subequations}
so that $E^\mu$ and $B^\mu$ can be called the {\em electric field
  4-vector} and {\em magnetic field 4-vector} respectively.   Of
course, it has to be acknowledged that the sign on the right side of
\Eqn{B0} depends on the choice of the components of the Levi-Civita
symbol.  Our convention has been described in the Appendix.  Note also
that
\begin{eqnarray}
  u^\mu E_\mu = 0 \,, \qquad u^\mu B_\mu = 0 \,,
  \label{uEuB}
\end{eqnarray}
relations which follow easily from \Eqn{EB} because of the
antisymmetry of $F^{\mu\nu}$ and $\tilde F^{\mu\nu}$.  

It seems that the two 4-vectors $E^\mu$ and $B^\mu$ contain all the
information that is there in $F^{\mu\nu}$.  That is indeed true.  If
fact, just as these 4-vectors can be determined from the field tensor,
the field tensor can also be reconstructed from these two 4-vectors.
The relation is~\cite{Padmanabhan:2008gr}
\begin{eqnarray}
  F^{\mu\nu} = E^\mu u^\nu - E^\nu u^\mu + 
  \varepsilon^{\mu\nu\lambda\rho} B_\lambda u_\rho \,.
  \label{FEB}
\end{eqnarray}
One can easily check, using \Eqn{uEuB}, that \Eqn{EB} follows from this
expression.  It is also instructive to write the Maxwell equations
using the electric field and the magnetic field 4-vectors.  Clearly,
contracting \Eqs{dF} {dFtil} with $u_\nu$, one obtains
\begin{subequations}
  \label{maxwellEB}
\begin{eqnarray}
  \del_\mu E^\mu &=& \frac1c J^\nu u_\nu \,, \\ 
  \del_\mu B^\mu &=& 0 \,.
\end{eqnarray}
The other two equations will read
\begin{eqnarray}
  \varepsilon^{\mu\nu\lambda\rho} u_\nu \del_\lambda E_\rho - 
  u^\alpha \del_\alpha B^\mu &=& 0 \,,
  \label{faradayEB} \\ 
  \varepsilon^{\mu\nu\lambda\rho} u_\nu \del_\lambda B_\rho +
  u^\alpha \del_\alpha E^\mu &=& - \frac1c (J^\mu - u^\mu u^\nu J_\nu)
  \,.
\end{eqnarray}
\end{subequations}

In the form given in \Eqn{FEB}, $F^{\mu\nu}$ is the sum of two
different antisymmetric tensors --- one composed of the electric
4-vector only, and the other composed of the magnetic 4-vector only.
We can now see how one might define an antisymmetric tensor containing
two constitutive parameters.  We can put two Lorentz-invariant
parameters to go with the two antisymmetric tensors to define
\begin{eqnarray}
  G^{\mu\nu} = \epsilon (E^\mu u^\nu - E^\nu u^\mu) +
  \frac1\mu \varepsilon^{\mu\nu\lambda\rho} B_\lambda u_\rho \,.
  \label{G}
\end{eqnarray}
Alternatively, we can define the 4-vectors
\begin{subequations}
  \label{DH}
\begin{eqnarray}
  D^\nu &=& \epsilon E^\nu \,, \\
  H^\nu &=& \frac1\mu B^\nu \,, 
\end{eqnarray}
\end{subequations}
and define 
\begin{eqnarray}
  G^{\mu\nu} = D^\mu u^\nu - D^\nu u^\mu +
  \varepsilon^{\mu\nu\lambda\rho} H_\lambda u_\rho \,.
  \label{GDH}
\end{eqnarray}
The two definitions are obviously equivalent.  The inhomogeneous
Maxwell equations involving the free sources is given by \Eqn{dG}.
The objects $\epsilon$ and $\mu$ are the {\em response functions}.

It is well-known that relations such as those in \Eqn{DH} are strictly
valid in terms of the Fourier transforms of the corresponding fields.
The point is that, the demand of linear response does not preclude
terms with extra derivatives in the definitions of $D^\nu$ and
$H^\nu$.  In the Fourier space, however, all these derivatives turn
into functions of the wave 4-vector $k^\mu$.  Thus, \Eqn{DH} really
says that the Fourier transform of 4-vectors $D^\nu$ and $H^\nu$ bear
linear relations to the Fourier transforms of $E^\nu$ and $B^\nu$, and
in each case the proportionality factor can be a function of $k^\mu$.

Earlier, we said that $\epsilon$ and $\mu$ are invariant.  Now we are
saying that they are functions of $k^\mu$.  Taken together, the two
statements mean that $\epsilon$ and $\mu$ can depend only on Lorentz
invariant quantities that can be constructed from the wave vector.
Using the medium 4-vector $u^\mu$, we see that there are two such
invariants, $k^\mu k_\mu$ and $k^\mu u_\mu$.  We can define the
independent variables to be
\begin{eqnarray}
  \omega = k^\mu u_\mu \,, \qquad K = \sqrt{(k\cdot u)^2 - k^\mu
    k_\mu} \,.
  \label{omK}
\end{eqnarray}
In the rest frame of the medium, $\omega$ becomes the frequency and
$K$ becomes the magnitude of the wave 3-vector.  Thus, the response
functions $\epsilon$ and $\mu$ can be functions of the frequency and
wavenumber, the latter being the magnitude of the 3-vector $\vec k$.
This is what is expected in an isotropic medium.  If the medium is not
isotropic, there will be other vectors associated with the medium, and
one will be able to construct more invariants.

\section{The Lagrangian}\label{s:lag}
We now want to see how we might obtain \Eqn{dG} from a Lagrangian.
Replacing $F^{\mu\nu}$ by $G^{\mu\nu}$ in \Eqn{lagF} would not do, for
the simple reason that it would yield Euler-Lagrange equations
containing quadratic combinations of the response functions.  However,
we can try
\begin{eqnarray}
  \lag = - \frac14 F_{\mu\nu} G^{\mu\nu} - \frac1c J^\mu \sub{free}
  A_\mu \,. 
  \label{lagG}
\end{eqnarray}
Note that \Eqn{ddA/ddA} gives
\begin{subequations}
  \label{derivEB}
\begin{eqnarray}
  {\del (E_\mu) \over \del (\del_\alpha A_\beta)} &=& \null -  
  \delta^\alpha_\mu u^\beta + \delta^\beta_\mu u^\alpha \,, \\ 
    {\del (B_\mu) \over \del (\del_\alpha A_\beta)} &=& - \null
    \varepsilon^{\mu\nu\alpha\beta} u_\nu   \,. 
\end{eqnarray}
\end{subequations}
Using the expression for $F^{\mu\nu}$ and $G^{\mu\nu}$ from \Eqs
{F}{G} and using the derivatives from \Eqn{derivEB}, it is easy to
verify that one obtains \Eqn{dG} as the Euler-Lagrange equation.

There is another way of looking at the Lagrangian density of
\Eqn{lagG} which might offer new insight.  Using \Eqs {uu}{uEuB}, one
finds
\begin{eqnarray}
  (u^\mu E^\nu - u^\nu E^\mu) (u_\mu E_\nu - u_\nu E_\mu) = 2 E^\nu
  E_\nu \,. 
\end{eqnarray}
Similarly, using \Eqn{app.2contr}, it is easy to show that
\begin{eqnarray}
  (\varepsilon^{\mu\nu\lambda\rho} u_\lambda B_\rho)
  (\varepsilon_{\mu\nu\alpha\beta} u^\alpha B^\beta) =
  - 2 B^\rho B_\rho \,,
\end{eqnarray}
whereas, because of the complete antisymmetry of the Levi-Civita
symbol, 
\begin{eqnarray}
  (u_\mu E_\nu - u_\nu E_\mu) \varepsilon^{\mu\nu\lambda\rho}
  u_\lambda B_\rho = 0 \,.
\end{eqnarray}
This exercise shows that we can write the Lagrangian density of
\Eqn{lagG} in the alternative form
\begin{eqnarray}
  \lag = - \frac12 \Big( \epsilon E^\alpha E_\alpha - \frac1\mu
  B^\alpha B_\alpha \Big) - \frac1c J^\mu \sub{free} A_\mu \,. 
  \label{LEB}
\end{eqnarray}
It is obvious that if we put $\epsilon=\mu=1$, then this expression
reduces to the Lagrangian density in the vacuum.  This is an
equivalent way of defining the response functions.

\section{Stress-energy-momentum tensor of the field}\label{s:sem}
The stress-energy-momentum tensor (or SEM tensor for the sake of
brevity) in the vacuum is given by
\begin{eqnarray}
  T^{\mu\nu} = \null - \eta_{\lambda\rho} F^{\mu\lambda} F^{\nu\rho} + 
  \frac14 \eta^{\mu\nu} F_{\lambda\rho} F^{\lambda\rho} \,.
  \label{Tvac}
\end{eqnarray}
In a medium, since we expect to obtain an expression that is linear in
the response functions, we should guess that the appropriate form
should be
\begin{eqnarray}
  T^{\mu\nu} = \null - \eta_{\lambda\rho} F^{\mu\lambda} G^{\nu\rho} +
  \frac14 \eta^{\mu\nu} F_{\lambda\rho} G^{\lambda\rho} \,.
  \label{T}
\end{eqnarray}
Recalling \Eqs{FEB}{GDH}, this result can be cast in the form
\begin{eqnarray}
  T^{\mu\nu} &=& \null - (E^\mu D^\nu + H^\mu B^\nu) - u^\mu u^\nu 
  \Big( E^\alpha D_\alpha + H^\alpha B_\alpha \Big)
   \next && - \Big(
  \varepsilon^{\mu\lambda\alpha\beta} u^\nu u_\alpha D_\lambda B_\beta
  + \varepsilon^{\nu\lambda\alpha\beta} u^\mu u_\alpha E_\lambda
  H_\beta \Big) + 
  \frac12 \eta^{\mu\nu} \Big( E^\alpha D_\alpha + H^\alpha B_\alpha
  \Big)  \,.
  \label{TEB}
\end{eqnarray}
It is easy to see that, in the rest frame of the medium, this tensor
has the components
\begin{subequations}
\begin{eqnarray}
  T^{00} &=& - \frac12 \Big( E^\alpha D_\alpha + H^\alpha B_\alpha
  \Big) = \frac12 \Big( \vec E \cdot \vec D + \vec H \cdot \vec B
  \Big) \,, \\
  T^{0i} &=& - \varepsilon^{ij0k} E_j H_k = \Big( \vec E \times \vec H \Big)^i \,, \\
  T^{i0} &=& - \varepsilon^{ij0k} D_j B_k = \Big( \vec D \times \vec B
  \Big)^i \,, \\ 
  T^{ij} &=& \null - (E^i  D^j  + H^i  B^j ) + 
  \frac12 \delta^{ij} \Big( \vec E \cdot \vec D + \vec H \cdot \vec B
  \Big) \,, 
\end{eqnarray}
\end{subequations}
which are the expected results, available in textbooks.

\section{Optical activity}\label{s:act}
Optical activity, and its connection with the Maxwell equations, is a
subject that is not discussed in most textbooks on Electrodynamics.
In one exception where it is discussed \cite{LL}, the property is
shown to be connected with the tensorial structure of the dielectric
function.  However, it was pointed out a while ago
\cite{Nieves:1988qz, Nieves:1992et} that it is not necessary to go
beyond the numerical response functions, as are appropriate for an
isotropic medium, in order to accommodate an explanation of natural
optical activity.  For this, it is important to realize that mimicking
the form for $F^{\mu\nu}$ given in \Eqn{F} to write down the
expression for $G^{\mu\nu}$ in \Eqn{G} does not give the most general
form for $G^{\mu\nu}$ for an isotropic medium.  It was shown in these
papers that the most general form for $G^{\mu\nu}$ should contain
another response function, which can explain the phenomenon of natural
optical activity.

In the covariant formulation that we have been discussing, there have
been several hints that something is not being considered.  For
example, if we look at \Eqn{LEB}, we see a term containing $E^\alpha
E_\alpha$ and a term containing $B^\alpha B_\alpha$, but no term
containing $E^\alpha B_\alpha$ \cite{Nieves:1989uh}.  Equivalently, we
can look at \Eqn{G} and ask ourselves, why have we not included terms
with the tensors
\begin{subequations}
  \label{extra}
\begin{eqnarray}
  \Gamma_E^{\mu\nu} &=& \varepsilon^{\mu\nu\lambda\rho} E_\lambda
  u_\rho \,, \label{extraE} \\
  \Gamma_B^{\mu\nu} &=& B^\mu u^\nu - B^\nu u^\mu \,, 
  \label{extraB}
\end{eqnarray}
\end{subequations}
which are obtained by interchanging the roles of the 4-vectors $E^\mu$
and $B^\mu$ in the two antisymmetric tensors that appear there?

It seems that there is a conflict between the two statements of
incompleteness that we just put forward.  In the Lagrangian
formulation, there seems to be just one term missing, and we can
rectify the incompleteness by writing
\begin{eqnarray}
  \lag = - \frac12 \Big( \epsilon E^\alpha E_\alpha - \frac1\mu
  B^\alpha B_\alpha - \zeta E^\alpha B_\alpha \Big) -
  \frac1c J^\mu \sub{free} A_\mu \,. 
  \label{Lzeta}
\end{eqnarray}
But if we look at \Eqn{extra}, it seems that it calls for two extra
coefficients to be introduced in the definition of $G^{\mu\nu}$, one
for each of the tensors defined there.  The resolution of this
apparent contradiction will be discussed shortly.

Meanwhile, let us proceed with \Eqn{Lzeta}.  The Euler-Lagrange
equation from this Lagrangian can be easily derived by using
\Eqn{derivs}, and the result is given by \Eqn{dG}, where
\begin{eqnarray}
  G^{\mu\nu} = \epsilon (E^\mu u^\nu - E^\nu u^\mu) +
  \frac1\mu \varepsilon^{\mu\nu\lambda\rho} B_\lambda u_\rho + \frac12
  \zeta \Big( B^\mu u^\nu - B^\nu u^\mu +
  \varepsilon^{\mu\nu\alpha\beta} E_\alpha u_\beta \Big) \,.
  \label{Gzeta}
\end{eqnarray}
As seen here, the expression that multiplies $\zeta$ is
$\Gamma_E^{\mu\nu} + \Gamma_B^{\mu\nu}$.  If two repsonse functions
were allowed corresponding to the two tensors shown in \Eqn{extra},
that would have meant that we could add another term proportional to
$\Gamma_E^{\mu\nu} - \Gamma_B^{\mu\nu}$ in this equation.  But,
starting from the definition of \Eqn{Ftil}, it is easy to see that
\Eqn{FEB} implies
\begin{eqnarray}
  \tilde F^{\mu\nu} = \null - B^\mu u^\nu + B^\nu u^\mu +
  \varepsilon^{\mu\nu\alpha\beta} E_\alpha u_\beta = \Gamma_E^{\mu\nu}
  - \Gamma_B^{\mu\nu} \,. 
  \label{FtilEB}
\end{eqnarray}
Adding a term proportional to it in the expression for $G^{\mu\nu}$
would not make any difference in the Euler-Lagrange equation, because
the dual tensor satisfies \Eqn{dFtil}.  Such a term would be
irrelevant.  This is why we can add only one extra response function,
not two.  Another way of saying it is that \Eqn{dFtil} means
\begin{eqnarray}
  \del_\mu \Gamma_E^{\mu\nu} = \del_\mu \Gamma_B^{\mu\nu} \,,
\end{eqnarray}
which is the same as \Eqn{faradayEB}.  Thus, we can use only one of
the two tensors $\Gamma_E^{\mu\nu}$ and $\Gamma_M^{\mu\nu}$ in the
expression for $G^{\mu\nu}$, augmenting the coefficient from
$\frac12\zeta$ to $\zeta$.

There is another interesting point about the $\zeta$-term.  If we take
\Eqn{app.5eps} given in the Appendix with $u^\mu$ in place of the
arbitary 4-vector, and contract with $F^{\mu\nu}F^{\lambda\rho}
u^\alpha$, we obtain the identity
\begin{eqnarray}
  E^\alpha B_\alpha = - \frac14 F^{\mu\nu} \tilde F_{\mu\nu} \,.
\end{eqnarray}
However, we also know that
\begin{eqnarray}
  F^{\mu\nu} \tilde F_{\mu\nu} &=& \frac12 \varepsilon_{\mu\nu\lambda\rho}
  F^{\mu\nu} F^{\lambda\rho} = 2 \varepsilon_{\mu\nu\lambda\rho}
  (\del^\mu A^\nu) (\del^\lambda A^\rho) \\
  &=& \del^\mu (2 \varepsilon_{\mu\nu\lambda\rho} A^\nu \del^\lambda
  A^\rho) \,,
\end{eqnarray}
a total derivative.  Such terms in a Lagrangian do not contribute to
the Euler-Lagrange equations.  

But obviously the $\zeta$-term makes a difference in the
Euler-Lagrange equation.  This apparent contradition implies that
$\zeta$ cannot be a constant.  In the momentum space, it should depend
on $\omega$ and $K$, introduced in \Eqn{omK}.  This conclusion was
reached from the non-relativistic treatment as well
\cite{Nieves:1992et}.

The effect of the $\zeta$-term can be easily understood by considering
the partiy transformation property of the different terms in either
the Lagrangian of \Eqn{Lzeta} or in the tensor defined in
\Eqn{Gzeta}.  It is easily seen that under parity,
\begin{subequations}
\begin{eqnarray}
  E^\mu &\xrightarrow{\quad{\rm parity}\quad}& P^\mu{}_\nu E^\nu \,,
  \\
  B^\mu &\xrightarrow{\quad{\rm parity}\quad}& - P^\mu{}_\nu B^\nu \,,
\end{eqnarray}
\end{subequations}
where
\begin{eqnarray}
  P^\mu{}_\nu = \mathop{\rm diag} \Big(+1,-1,-1,-1 \Big) \,.
\end{eqnarray}
Thus, the terms involving $\epsilon$ and $\mu$ are parity invariant,
whereas the terms involving $\zeta$ are not.  Because of this, the
presence of a non-zero value of $\zeta$ implies different dispersion
relations for the right-circular and left-circular polarizations of
electromagnetic waves in the medium \cite{Nieves:1988qz,
  Nieves:1992et}, which is the root cause for optical activity.

Finally, we might wonder about the form of the SEM tensor in presence
of optical activity.  Since $G^{\mu\nu}$ contains the activity
constant $\zeta$, it is naively expected that when this expression in
substituted into \Eqn{T}, we will see some $\zeta$-dependence in the
SEM tensor.  However, once the substitution is made, it is seen that
the $\zeta$-dependent terms all cancel out and we obtain exactly
\Eqn{TEB}.

\paragraph*{Acknowledgements~:}  I thank Kaushik Bhattacharya for
discussions.  The work was supported by the SERB research grant
EMR/2017/001434 of the Government of India.

\appendix

\setcounter{equation}0
\renewcommand{\theequation}{A.\arabic{equation}}

\section*{Appendix}\label{s:app}
Here we collect a few results related to the 4-dimensional Levi-Civita
symbol.  We define it as
\begin{eqnarray}
  \varepsilon_{\mu\nu\lambda\rho} = \begin{cases}
    +1 & \mbox{if the indices form an even permutation of
      0,1,2,3, } \\
    -1 & \mbox{if the indices form an even permutation of
      0,1,2,3, } \\
    0 & \mbox{otherwise.}
  \end{cases}
  \label{app.epsdn}
\end{eqnarray}
The usual rules of raising and lowering indices would then imply 
\begin{eqnarray}
  \varepsilon^{\mu\nu\lambda\rho} = \begin{cases}
    -1 & \mbox{if the indices form an even permutation of
      0,1,2,3, } \\
    +1 & \mbox{if the indices form an even permutation of
      0,1,2,3, } \\
    0 & \mbox{otherwise.}
  \end{cases}
  \label{app.epsup}
\end{eqnarray}
We have used the expression obtained by contracting two indices of a
pair of Leve-Civita symbols:
\begin{eqnarray}
  \varepsilon^{\mu\nu\lambda\rho} \varepsilon_{\mu\nu\alpha\beta} = -
  2(\delta^\lambda_\alpha \delta^\rho_\beta -
  \delta^\lambda_\beta \delta^\rho_\alpha) \,.
  \label{app.2contr}
\end{eqnarray}

Another important identity, used in the text, involves an arbitrary
4-vector $V^\mu$ and states that
\begin{eqnarray}
  \varepsilon^{\mu\nu\lambda\rho} V^\alpha -
  \varepsilon^{\alpha\nu\lambda\rho} V^\mu -
  \varepsilon^{\mu\alpha\lambda\rho} V^\nu -
  \varepsilon^{\mu\nu\alpha\rho} V^\lambda -
  \varepsilon^{\mu\nu\lambda\alpha} V^\rho = 0 \,.
  \label{app.5eps}
\end{eqnarray}
The proof is very simple.  It can be checked easily that the left side
of the equation is completely antisymmetric in the interchange of any
pair of indices, and is therefore a rank-5 antisymmetric tensor.
Since five indices cannot be antisymmetrized in a 4-dimensional
geometry, the left side must vanish.

\bibliographystyle{unsrt}
\bibliography{covem.bib}

\end{document}